\documentclass[preprint]{aastex}

\newcommand{\kic}{KIC~8462852}

\slugcomment{Published on ApJL, 814, L15}

\shorttitle{KIC~8462852 Infrared Flux}
\shortauthors{Marengo et al. 2015, ApJL, 814, L15}

\begin{document}


\title{KIC~8462852 -- The Infrared Flux}


\author{Massimo Marengo and Alan Hulsebus}
\affil{Department of Physics and Astronomy, Iowa State University, Ames, IA 50011}

\and

\author{Sarah Willis\altaffilmark{1}}
\affil{Harvard-Smithsonian Center for Astrophysics, Cambridge, MA 02138}

\altaffiltext{1}{Current affiliation: MIT Lincoln Laboratory, Cambridge, MA, USA}



\begin{abstract}
We analyzed the warm Spitzer/IRAC data of \kic. We found no evidence of infrared excess at 3.6~\micron{} and a small excess of $0.43 \pm 0.18$~mJy at 4.5~\micron, below the $3\sigma$ threshold necessary to claim a detection. The lack of strong infrared excess 2 years after the events responsible for the unusual light curve observed by Kepler, further disfavors the scenarios involving a catastrophic collision in a \kic{} asteroid belt, a giant impact disrupting a planet in the system or a population of a dust-enshrouded planetesimals. The scenario invoking the fragmentation of a family of comets on a highly elliptical orbit is instead consistent with the lack of strong infrared excess found by our analysis.
\end{abstract}


\keywords{stars: individual (KIC~8462852) --- infrared: stars}



\section{Introduction}\label{s:introduction}

\objectname[KIC 8462852]{\kic}, also known as \objectname[TYC 3162-665-1]{TYC~3162$-$665$-$1} and \objectname[2MASS J20061546+4427248]{2MASS~J20061546$+$4427248}, is a $V \simeq 12$~mag star in the field of the Kepler space telescope primary mission \citep{2010Sci...327..977B}. This star was identified serendipitously by the Planet Hunters project \citep{2012MNRAS.419.2900F} for its unusual light curve, characterized by deep dimming (down to below $\sim 20$\% of the stellar flux) lasting between 5 and 80 days and with an irregular cadence and unusual profile (\citealt{2015arXiv150903622B}, B15 hereafter). Since its discovery, \kic{} has been the subject of intense multi-wavelength monitoring and has spurred wild speculations about the nature of the bodies, or structures, responsible for the dimming of its visible flux (see, e.g. \citealt{2015arXiv151004606W}).

The star has been carefully characterized in B15. High-resolution ($R \sim 47000$) spectroscopic observations obtained with the FIES spectrograph at the Nordic Optical Telescope in La Palma, Spain, revealed that \kic{} is a main-sequence star with an effective temperature $T_{eff} = 6750 \pm 120$~K, $\log g = 4.0 \pm 0.2$ and solar metallicity consistent with an F3V star. The spectral energy distribution (SED) of the source, obtained combining ground-based $BV(RI)_c$ and 2MASS \citep{2006AJ....131.1163S} $JHK_s$ photometry, with space-based NUV Galex \citep{2007ApJS..173..682M} and mid-IR WISE \citep{2010AJ....140.1868W} data, is also consistent with the spectroscopic identification of the source. Of particular significance is the lack of measurable infrared excess in the WISE photometry. Careful fitting of the star's SED with a stellar atmosphere model revealed that the source is located at a distance of 454~pc, with a reddening of $E(B-V) = 0.11$~mag.

Natural guide star adaptive optics (AO) imaging obtained at the Keck II telescope on Mauna Kea, Hawaii, using the NIRC2 infrared ($J$, $H$, and $K$ bands) camera revealed the presence of a fainter source 1.95\arcsec{} East from \kic. The brightness ratio of this source with respect to the primary ($\sim 2$\%), as well as its near-IR colors, is consistent with this source being an M2V companion to \kic. While the physical association of the two stars cannot be demonstrated with current data, B15 estimated that a chance alignment between the two sources would be only $\sim 1$\%. If the pair is indeed a physical binary, at the distance of \kic{} their separation would correspond to $885 \ AU$.

As mentioned before, \kic{}'s claim to fame is related to the very unusual dips in its Kepler time-series photometry. One event near JD~2455626 (2011 March 5) lasting for $\sim 3$~days and a series of events starting from JD~2456343 (2013 February 19) lasting for $\sim 60$~days stand out for their unusual shape, lack of periodicity, and depth. A search through the Kepler database for targets with similar dips turned came up empty. Realistic scenarios for the phenomena observed for this star are discussed at length in B15. They are related to the episodic occultation of the star by a circumstellar dust clump, either produced in the aftermath of a catastrophic collision in the system's asteroid belt or a giant impact in the system or associated with a population of dust-enshrouded planetesimals, or produced by the breakout of a family of comets. The parameter space for all these scenarios is constrained by the lack of infrared excess above the expected stellar photospheric emission in the WISE photometry. For this reason, the breakout of a family of comets remains the most likely possibility, as it requires the least amount of mid-infrared excess.

The WISE observations of \kic, however, were restricted to Kepler Q5, nine months before the first dimming event in 2011 March (Q8). The possibility remains that infrared observations of the system, if carried out \emph{after} the dimming events, would be capable of detecting some lingering excess. The star was in fact observed by the IRAC camera \citep{2004ApJS..154...10F} on board the Spitzer Space Telescope \citep{2004ApJS..154....1W} in early 2015 as part of the SpiKeS program aimed to map the entire Kepler field.

In this Letter we present the Spitzer/IRAC photometry of \kic. In section~\ref{s:obs}, we describe the observations and how we measured the photometry of the source. In section~\ref{s:excess}, we analyze the results in order to assess the presence of an infrared excess, taking into account the added flux contributed by the nearby M2 star. The consequences of our measurement on the standing hypothesis for the nature of the dimming events are discussed in section~\ref{s:disc}; our results are summarized in section~\ref{s:sum}.

\section{Spitzer/IRAC Observations and Photometry}\label{s:obs}

\kic{} was observed on 2015 January 18 (JD~2457040.87) as part of the Spitzer Kepler Survey (SpiKeS; PID~10067, PI M. Werner) aimed at mapping the entire Kepler field with warm IRAC. The program uses a shallow mapping strategy consisting of overlapping frames, with the standard 12~sec frame time Astronomical Observation Template (equivalent to a 10.4~sec exposure time), organized in such a way that each point in the sky is observed in at least three consecutive frames (for a total of 31.2~sec) in both warm IRAC bands (3.6 and 4.5~\micron). \kic{} was observed as part of AOR~52340480. The Basic Calibrated Data (BCD) from the observation are publicly available from the Spitzer Heritage Archive and have been processed by the Spitzer pipeline, version S19.1.0. 

\begin{figure}
\includegraphics[angle=0,scale=0.90]{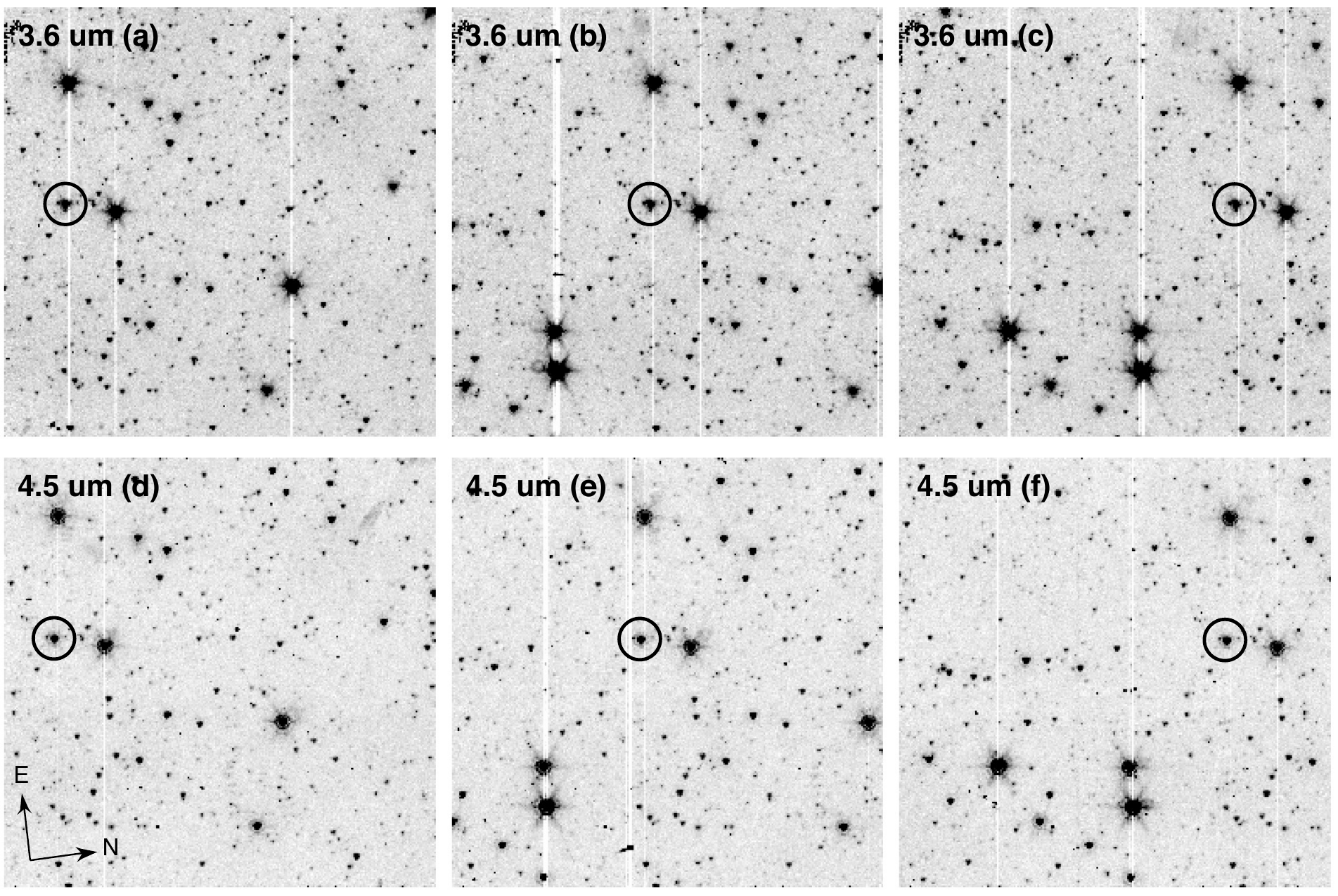}
\caption{IRAC BCDs containing \kic{} (3.6~\micron{} top row, 4.5~\micron{}, bottom row). The target is marked with a circle (15\arcsec{} radius). Note the significant vertical lines (column pulldown artifacts) caused by the nearby source KIC~8462934, especially strong at 3.6~\micron. One 4.5~\micron{} frame (panel e) is also affected by even stronger pulldown artifacts from cosmic rays at the bottom of the frame. The images are in array coordinates (position angle 82.66$^\circ$).}\label{f:bcds}
\end{figure}

\kic{} is easily identified based on its coordinates, provided in B15. Figure~\ref{f:bcds} shows the six BCDs (three for each IRAC channel) containing the source. Given the crowding of Kepler's field, the images are heavily affected by vertical electronic artifacts caused by saturated sources (known as \emph{column pulldown}). One of these artifacts (due to the nearby $V \simeq 11.5$~mag source KIC~8462934) affects \kic{} in all frames. One of the 4.5~\micron{} frames is also affected by two more pulldown artifacts from cosmic rays at the bottom of the frame. These artifacts need to be removed as they can affect the photometry of the target. For this reason in the following analysis, we have used the artifact-corrected frames (CBCDs) where the offset in the columns affected by pulldown is determined separately above and below the triggering source and corrected.

We have then determined the magnitude of \kic{} following the prescriptions in the Spitzer Science Center 2013 July 18 memo.\footnote{http://ssc.spitzer.caltech.edu/warmmission/news/18jul2013memo.pdf} We performed photometry separately on each individual CBCD, with the aperture position determined with flux-weighted moments, and then applied corrections for array location-dependent and pixel-phase photometry. The main source of photometric uncertainty with this method is related to IRAC's point-spread function (PSF) undersampling, causing small photometric variations from frame to frame. To take this issue into account, we have estimated the photometric uncertainty as the standard deviation of our three measurements, corrected for bias (by a factor of 1/0.866 appropriate for $n = 3$).

We adopted an aperture of 3 IRAC pixel radius (corresponding to 3.65\arcsec) with a sky annulus of 3 to 7 IRAC pixels radius for background subtraction. The aperture corrections for such a combination of aperture and sky annulus (the same used for the IRAC absolute calibration) is equal to 0.1286 at 3.6~\micron{} and 0.1256 at 4.5~\micron{} \citep{2012SPIE.8442E..1ZC}. The conversion from instrumental to Vega magnitudes was done relying on IRAC's absolute photometric calibration (using the warm mission flux conversion factors of 0.1253 and 0.1469 MJy/sr per DN/sec and the Vega fluxes of 280.9 and 179.7~Jy at 3.6 and 4.5~\micron, respectively).

The IRAC magnitudes we obtained with the process outlined above are listed in Table~\ref{t:mags}, together with the 2MASS and WISE ($W1$ and $W2$ bands only) photometry for comparison. The magnitudes listed in Table~\ref{t:mags} are before color correction and do not include the uncertainty due to the instrument absolute photometric calibration. As shown in \citet{2012SPIE.8442E..1ZC} the uncertainty in the IRAC absolute calibration can be split in two parts. One is the statistical error in the IRAC flux conversion factors, equal to 0.6\% and 0.5\% at 3.6 and 4.5~\micron{} respectively. We have included this uncertainty in the analysis described in the next section. The other is the error in the calibration zero points, which can be as high as 1.5\%. It should be noted, however, that the IRAC calibration zero point is based on observed spectra and models of a set of calibrators whose absolute normalization relied on their 2MASS magnitudes \citep{2005PASP..117..978R, 2003AJ....126.1090C}. The absolute calibration of WISE is similarly tied to the one of Spitzer \citep{2010AJ....140.1868W, 2011ApJ...735..112J}. This was done to ensure that the comparison between IRAC, 2MASS and WISE fluxes is not affected by systematic errors to the same degree as their absolute calibration zero points. A test performed comparing the IRAC photometry of a set of calibrators to the HST CALSPEC database indeed found no systematic bias \citep{2012SPIE.8442E..1ZC}.

The IRAC magnitudes of \kic{} are consistent with the corresponding WISE photometry, despite the small differences in the transmission curves of the photometric systems adopted in the two instruments. The photometric uncertainty of the IRAC measurements at 3.6 and 4.5~\micron{}, however, are about 4 and 2 times smaller than the corresponding WISE magnitude uncertainties.

\section{Infrared Excess}\label{s:excess}

To search for evidence of an IRAC infrared excess in \kic, we performed a $\chi^2$ fit of the star's photometry with available stellar atmosphere models. We have derived synthetic photometry for each passband by convolving the model spectra with the appropriate transmission curve and then compared them with the measured in-band fluxes after applying color corrections.

We determined the normalization of the models by fitting the 2MASS near-IR photometry, rather than the whole optical and near-IR SED (as done by B15) in order to limit the effects of the systematic uncertainties in the cross calibration between optical and infrared data. While this choice could partially alias an infrared excess in the IRAC bands if a near-IR excess is also present, this is an unlikely scenario as it would require very hot dust close to the star, which is rejected by the analysis in B15. Furthermore, the 2MASS magnitudes were measured many years before the Kepler dimming events: finding an excess in the 2MASS observation would imply that the \kic{} phenomenon existed for a long time before being discovered by Kepler.

Before normalizing the stellar models with the 2MASS photometry, however, we needed to perform two different sets of corrections. The first is to correct the magnitudes in Table~\ref{t:mags} for the $E(B-V) = 0.11$ reddening measured by B15. We adopted the \citet{1989ApJ...345..245C} reddening law for 2MASS, and the total-to-selective extinction reported in \citet{2012ApJ...759..146M} for the two IRAC bands. The second correction requires the subtraction of the flux from the M dwarf companion found with the Keck/AO imaging. Regardless of the physical association with the primary, this red source is close enough to the primary to be blended in the IRAC, 2MASS and WISE photometry. To separate the contribution of the M dwarf companion from the total flux measured by 2MASS we used the accurate $J$, $H$ and $K$ magnitude differences reported by B15. Assuming that the source is indeed an M2V companion, we extrapolated its flux to the IRAC bands from the 2MASS flux in the $K$ band. For this task, we used Spitzer's flux estimator for stellar point sources (STAR-PET\footnote{http://ssc.spitzer.caltech.edu/warmmission/propkit/pet/starpet/index.html}) for an M2V spectral type. The fluxes of the corrected \kic{} primary photometry are listed in Table~\ref{t:excess}.

We started with an ATLAS9 model grid \citep{2004astro.ph..5087C}, which is the same used for the IRAC and 2MASS absolute calibration \citep{2012SPIE.8442E..1ZC}, setting gravity and metallicity to the values measured by B15 ($\log g = 4.0$ and $\log [Z/\textrm{H}] = 0.00$). The best fit temperature for the primary was determined by minimizing the $\chi^2$ of quadratically interpolated models with temperatures between 6750 and 7250~K. The best fit was obtained for $T_{eff} = 6950$~K ($\chi^2 = 0.36$) close to the spectroscopically determined effective temperature ($T_{eff} = 6750 \pm 120$~K, yielding instead $\chi^2 = 1.42$). Since the 2MASS photometry of \kic{} has a relative error of 0.020~mag (or $\sim 2.00$\%), we attributed to the atmosphere model normalization factor an uncertainty of 2\%$/ \sqrt{3} \simeq 1.13$\%. The good agreement of the 2MASS photometry with the best fit model supports the absence of a detectable IR excess in the 2MASS data.

\begin{figure}
\includegraphics[angle=-90,scale=0.70]{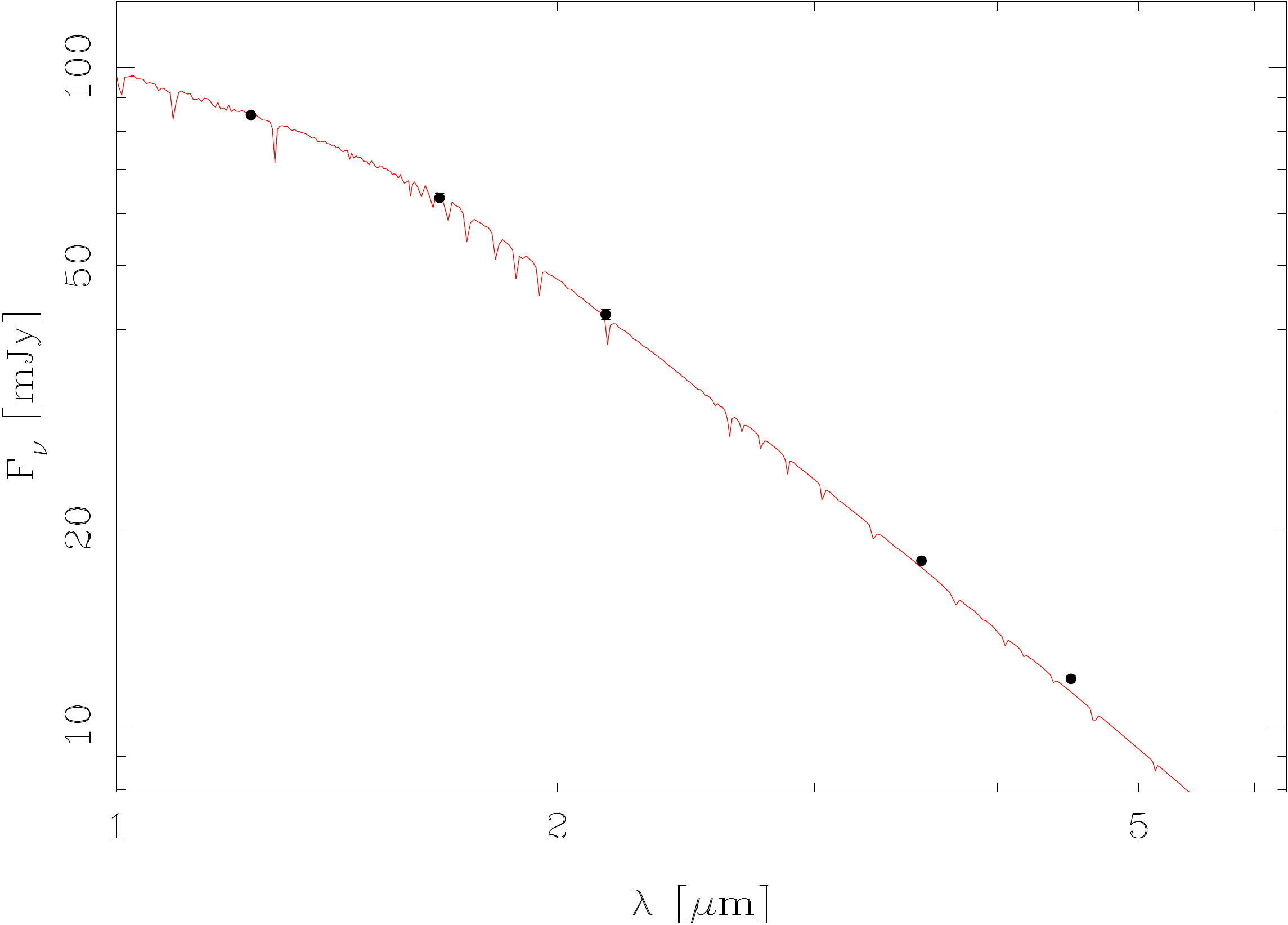}
\caption{Our reference ATLAS9 model atmosphere for an F3V star normalized with the 2MASS photometry. IRAC photometry shows a $\sim 2.4 \sigma$ excess at 4.5~\micron. Error bars for the IRAC bands are smaller than the symbols used in the figure.}\label{f:sed}
\end{figure}

The fluxes of the ATLAS9 atmosphere described above ("reference model" hereafter) are listed in Table~\ref{t:excess}, and the synthetic spectrum (together with the 2MASS and IRAC photometry) is shown in Figure~\ref{f:sed}. The IRAC photometry of the \kic{} primary is consistent with no excess at 3.6~\micron. A small excess of $0.43 \pm 0.18$~mJy (corresponding to a significance of $2.4\sigma$) is found above the photosphere model at 4.5~\micron. Letting $\log g$ and $\log [Z/\textrm{H}]$ vary within their measurement uncertainty has a small effect on the 4.5~\micron{} excess, and its significance remains in the 2.2 to $2.5\sigma$ range. To test the effect of using different model atmospheres, we repeated the same exercise using the PHOENIX grid \citep{2005ESASP.576..565B}. These models tend to have larger mid-IR emission, as suggested by \citet{2010MNRAS.409L..49S}, lowering the significance of the 4.5~\micron{} excess to the 1.8 to $2.1\sigma$ range.

The low significance of our measurement prevents us from concluding that the fluxes in Table~\ref{t:excess} represent a detection of mid-IR excess from \kic. They are nevertheless suggesting that in 2015 January this excess could have been present at 4.5~\micron, where its significance is above $2\sigma$ for most models. In absence of a firm detection, however, we set $3\sigma$ limits in the \kic{} excess equal to 0.75~mJy at 3.6~\micron{} and 0.54~mJy at 4.5~\micron{}.

\section{Discussion}\label{s:disc}

\begin{figure}
\includegraphics[angle=-90,scale=0.70]{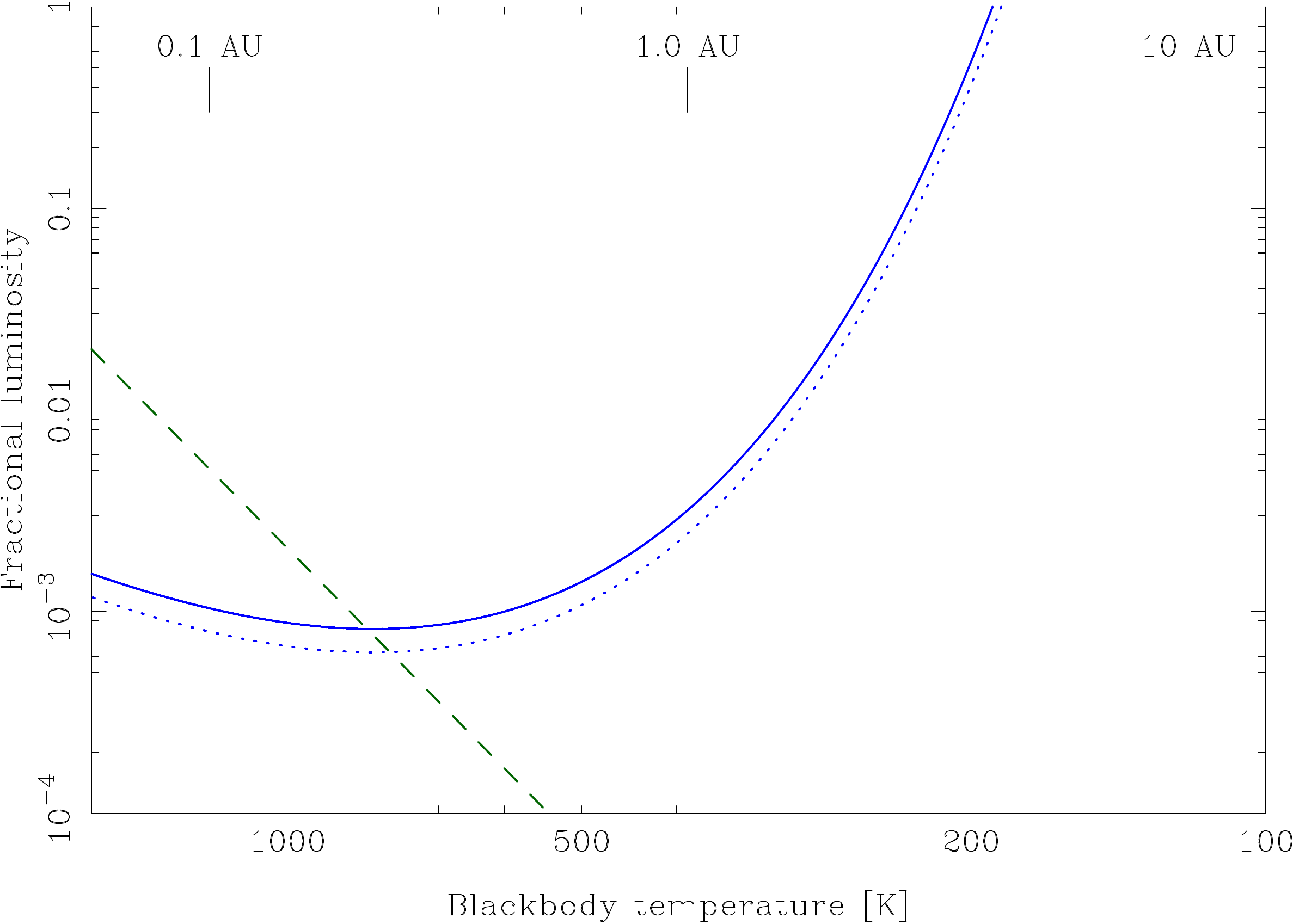}
\caption{Fractional luminosity of a clump of dust in proximity to \kic{} as a function of its blackbody temperature. The dotted line corresponds to the $2.4\sigma$ excess derived at 4.5~\micron{} with our reference Kurucz model. The solid line is instead the upper limit based on our $3\sigma$ limit for the excess in the 4.5~\micron{} IRAC band. The green dashed line is the dust fractional luminosity estimated in B15 by integrating the optical depth  in the Kepler light curve, assuming that the clumps are similar in size to the star and are on a circular orbit. The distances at the top of the figure are the reference blackbody orbital radii from B15.}\label{f:tbb}
\end{figure}

A mid-IR detection of excess emission from \kic{} (or a detection limit) would provide strong constraints on the nature and the location of the objects responsible for the dimming of the star. In particular, the fractional brightness of the excess listed in Table~\ref{t:excess} at 4.5~\micron{} ($f_{4.5} = F^{dust}_{4.5}/F^{star}_{4.5} \simeq 4.6 \times 10^{-2}$) can be translated into a total fractional luminosity $f = L_{dust}/L_{star}$, as a function of the dust temperature $T_d$. Following B15, this relation is shown in Figure~\ref{f:tbb} (dotted line). The solid line is instead the upper limit in the fractional luminosity based on our $3\sigma$ upper limit in the excess. The dashed line on the plot represents the relation between the fractional luminosity and the dust temperature estimated by B15 by integrating the optical depth inferred from the Kepler light curve. This estimate  assumes that the dust clumps responsible for the dimming are similar in size to the star and are on a circular orbit. Based on these hypotheses, our $3\sigma$ limit in the excess corresponds to a maximum fractional luminosity $f \approx 8 \times 10^{-4}$, a maximum dust blackbody temperature $T_d \approx 800$~K and to an orbital radius larger than $\approx 0.2 \ AU$. 

The absence of strong infrared excess at the time of the IRAC observations (after the dimming events) implied by our 4.5~\micron{} $3\sigma$ limit suggests that the phenomenon observed by Kepler produced a very small amount of dust. Alternatively, if significant quantity of dust is present, it must be located at large distance from the star. As noted by B15, this makes the scenarios very unlikely in which the dimming events are caused by a catastrophic collision in \kic{} asteroid belt, a giant impact disrupting a planet in the system, or a population of dust-enshrouded planetesimals. All these scenarios would produce very large amount of dust dispersed along the orbits of the debris, resulting in more mid-IR emission than what can be inferred from the optical depth of the dust seen passing along our line of sight to the star. Our limit (two times lower than the limit based on WISE data) further reduces the odds for these scenarios.

The giant impact hypothesis would be entirely ruled out \emph{if} the $2.4\sigma$ excess from our reference model is real and \emph{if} all dust transited the star during the Kepler dimming event, since in these hypotheses, Figure~\ref{f:tbb} would imply that the impact must have happened just beyond $0.2 \ AU$ from the star. This would be in contradiction with the requirement, based on the time lag between the two dimming events in the Kepler light curve, that the impacted body had an orbital radius of $\approx 1.6 \ AU$. If these hypotheses are not verified, a giant impact could still be possible at radii below our solid line and to the right of the dashed line in Figure~\ref{f:tbb}.

The hypothesis of the disruption of a family of comets is the preferred scenario in B15 because it dispenses with the requirement of a circular orbit, allowing the cloud of dust produced in the comets' fragmentation to rapidly move away from the star on a highly elliptical orbit. It remains the most likely hypothesis even with our 4.5~\micron{} $3\sigma$ limit since the 2 year gap between the Kepler events and the IRAC observations would have been sufficient for the cometary debris to move several $AU$ away from the tidal destruction radius of the star. At such a distance, the thermal emission from the dust would be peaked at longer wavelengths and undetectable by IRAC. A robust detection at longer wavelengths (where the fractional brightness of the debris with respect to the star would be more favorable) will allow the determination of the distance of the cometary fragments and constrain the geometry of this scenario.

\section{Summary and Conclusions}\label{s:sum}

We have analyzed the warm Spitzer/IRAC images of \kic{} obtained in 2015 January 18 as part of the SpiKeS survey of the Kepler field. Our mid-IR  photometry is in agreement with the photospheric emission from the star at 3.6~\micron{} and shows a small excess of $0.43 \pm 0.18$~mJy at 4.5~\micron, below the required $3\sigma$ confidence level for a detection. This reveals that 2 years after the dimming event observed by Kepler, no significant amount of circumstellar dust can still be detected. This further reduces the odds that the phenomena observed in 2011 and 2013 are caused by catastrophic collisions in the asteroid belt of this star, by a giant impact on one of its planets, or by a population of dust-enshrouded planetesimals because such extreme systems (e.g. BD$+20$ 307) do present a large excess detected in the 3-5~\micron{} spectral region (see e.g. \citealt{2015ApJ...805...77M} and references therein). The scenario in which the dimming in the \kic{} light curve were caused by the destruction of a family of comets remains the preferred explanation for the undetectable amount of infrared excess associated with the Kepler events.

Our measurement is not sufficient to constrain the temperature and the magnitude of the fractional luminosity of the dust that would be associated with the comets. However, if combined with detections at longer wavelengths (where the fractional infrared brightness is expected to be higher) and long-term infrared monitoring, it will allow constraining the temperature and location of the dust cloud and possibly the geometry of the catastrophic event at the root of this unusual phenomenon.



\acknowledgments

This work is based in part on observations made with the Spitzer Space telescope, which is operated by the Jet Propulsion Laboratory, California Institute of Technology, under a contract with NASA. We would like to thank the anonymous referee for his/her suggestions that significantly helped improve this work.



{\it Facilities:} \facility{Spitzer (IRAC), Kepler}



\



\clearpage





\clearpage

\clearpage

\clearpage







\begin{deluxetable}{ccl}
\tablecaption{\kic{} Infrared Magnitudes \label{t:mags}}
\tablewidth{0pt}
\tablehead{
\colhead{Band} & \colhead{Magnitude} & \colhead{Reference}
}
\startdata
$J$      & $10.763 \pm 0.021$      & 2MASS            \\
$H$     & $10.551 \pm 0.019$      & 2MASS             \\
$K_s$  & $10.499 \pm 0.020$      & 2MASS             \\
$[3.6]$ & $10.4768 \pm 0.0059$  & IRAC                 \\
$[4.5]$ & $10.4374 \pm 0.0107$  & IRAC                  \\
$W1$   & $10.425 \pm 0.023$      & (ALL) WISE        \\
$W2$   & $10.436 \pm 0.020$      & (ALL) WISE        \\
\enddata
\end{deluxetable}

\clearpage

\begin{deluxetable}{cccc}
\tablecaption{Infrared Excess (Reference Model)\label{t:excess}}
\tablewidth{0pt}
\tablehead{
\colhead{Band} & \colhead{$F_\nu$ (Model)} & \colhead{$F_\nu$ (Primary)} & \colhead{Excess }\\
\colhead{} & \colhead{[mJy]} & \colhead{[mJy]} & \colhead{[mJy]}
}
\startdata
$J$      & $84.61 \pm 0.98$ & $84.65 \pm 1.53$ &  \\ 
$H$     & $63.79 \pm 0.74$ & $63.36 \pm 1.08$ &  \\ 
$K_s$  & $41.85 \pm 0.48$ & $42.17 \pm 0.78$ &  \\ 
$[3.6]$ & $17.74 \pm 0.20$ & $17.82 \pm 0.15$ & $0.09 \pm 0.25$ \\ 
$[4.5]$ & $11.37 \pm 0.13$ & $11.79 \pm 0.13$ & $0.43 \pm 0.18$ \\ 
\enddata
\end{deluxetable}

\clearpage








\end{document}